\documentclass[12pt]{iopart}

\usepackage{graphicx}                
\usepackage[superscript,biblabel]{cite}
\usepackage{amssymb} 
\usepackage{mhchem}
\usepackage{here}                    
\usepackage{subcaption}              
\usepackage{url}                     
\usepackage{enumitem}	
\usepackage[unicode,colorlinks=true,allcolors=blue]{hyperref} 
\usepackage[capitalize,noabbrev]{cleveref}    
\usepackage{siunitx}
\usepackage{bm}
\usepackage{ulem}
\DeclareUnicodeCharacter{2009}{ }

\newcommand{\revone}[1]{#1}
\newcommand{\revonedel}[1]{}

\begin{document}

\title[QuSpin Zero-Field Magnetometer Characterization for the TUCAN Experiment]{QuSpin Zero-Field Magnetometer Characterization for the TUCAN Experiment}

\author{M~Zhao$^{1,2,}$\footnote{Corresponding Author}, R~Mammei$^{3}$, and D~Fujimoto$^1$}

\address{$^1$TRIUMF, 4004 Wesbrook Mall, Vancouver, BC V6T 2A3, Canada}
\address{$^2$Department of Physics and Astronomy, University of British Columbia, Vancouver, BC V6T 1Z1, Canada}
\address{$^3$Department of Physics and Astronomy, University of Winnipeg, Winnipeg, MB R3B 2E9, Canada}
\ead{mzhao17@student.ubc.ca}
\vspace{10pt}
\begin{indented}
\item[]August 2024
\end{indented}

\begin{abstract}
The TUCAN nEDM experiment \revonedel{characterizes}\revone{utilizes} the QuSpin Zero-Field Magnetometer (QZFM) to accurately map residual fields within a large magnetically shielded room. \revone{Three potential flaws of the QZFM are characterized in preparation for mapping.} The magnetometer's intrinsic offset was measured to be within \SI{\pm 3}{nT} and stable over a period of one year. The response was shown to be within 2 percent of linearity in the zero-field regime, up to \SI{2}{nT_{pp}}, and then follows a smooth dispersion curve. Crosstalk effects induced by multisensor operation were determined to have a small effect, and inconsequential with a separation above \SI{6}{cm}. These results enable the QZFM for accurate measurement of DC fields, increase the operational range of QZFM by a factor of more than an order of magnitude, and allow for higher efficiency and flexibility by green-lighting simultaneous operation of multiple QZFMs. 
\end{abstract}

%
\vspace{2pc}

\submitto{\MST}

 
\ioptwocol

\section{Introduction}
\revone{The asymmetry of matter over antimatter in the observed universe is a longstanding problem in physics. Sakharov's conditions for such an asymmetry require the existence of charge-parity (CP) symmetry violation in the standard model of particle physics \cite{sakharov_violation_1991}. While a small degree of CP violation can be found in the electroweak sector, additional mechanisms which break CP symmetry are needed to account for the quantity of matter we observe. If one were able to measure a permanent non-zero electric dipole moment (EDM) in a subatomic system, it would be direct evidence for new CP-violating physics. This would be immensely important for solving this issue.} The TRIUMF Ultracold Advanced Neutron (TUCAN) Collaboration aims to measure the neutron electric dipole moment (nEDM), $d_n$, to a target sensitivity of \SI{e-27}{\elementarycharge~\cdot~ cm}, a precision level that surpasses the current limit of $|d_n|<$\SI{1.8e-26}{\elementarycharge~\cdot~ cm} by an order of magnitude \cite{abel_measurement_2020}. The measurement of $d_n$ from ultracold neutrons (UCNs) is based on their spin precession frequency,  $\omega$, in a static magnetic field. A non-zero nEDM modifies the precession frequency in the presence of an electric field of magnitude $E$. The nEDM can be expressed as
\begin{equation}
\label{eq:d_n}
    d_n = \frac{\hbar(\omega_{\uparrow\uparrow}-\omega_{\uparrow\downarrow})}{4E},
\end{equation}
where $\hbar$ is the reduced Plank's constant, and $\uparrow\uparrow$, $\uparrow\downarrow$ denote parallel and anti-parallel configurations of the electric and magnetic fields. In contrast to competing nEDM experiments \cite{ayres_design_2021,nedm_collaboration_nedm_2014,ito_performance_2018}, the TUCAN experiment distinguishes itself in anticipating a significantly higher UCN count \cite{sidhu_estimated_2023, ahmed_first_2019, schreyer_optimizing_2020, sidhu_improving_2023}. Nevertheless, magnetometry remains important for reducing systematic effects.

To reduce the influence of stray fields in the experimental area, a 5-layer magnetically shielded room (MSR) was constructed to suppress field fluctuations to sub-\unit{pT} levels in the spin precession volume. Optically pumped magnetometers (OPMs) have become popular for monitoring fields in these ultra-quiet environments because of their ability to reach the \unit{fT} sensitivity range \revone{with a small and versatile form factor, a requirement not met by competing magnetometer technologies such as fluxgates, super-conducting quantum interference devices, and diamond magnetometers.} While the TUCAN experiment will correct for residual fields with purpose-built scalar OPMs \cite{klassen_magnetically_2024, klassen_all-optical_2020}, commercially available vector OPMs are useful for assessing the MSR performance, including its shielding factor, smoothness of internal gradients, and the effect of various degaussing protocols \cite{ayres_achieving_2024, ayres_very_2022}. The TUCAN collaboration has two 3rd generation tri-axial QuSpin Zero-Field Magnetometers (QZFMs) \cite{noauthor_quspin_nodate} in its inventory. In characterizing the sensors, the authors are primarily concerned with internal offsets when measuring absolute DC fields, the response linearity, and multisensor crosstalk. 

Published calibrations of QZFM sensors have demonstrated response non-linearity under background field strengths up to \SI{3}{nT_{pp}} and observed small crosstalk effects in a helmet-based sensor configuration for magnetoencephagraphy \cite{boto_moving_2018}. Tierney et al introduced a mathematical model for crosstalk-induced gain change based on the way mutual interference of sensor modulation fields modifies the OPM signal equation \cite{tierney_optically_2019}. In the special case where the modulation fields are parallel, the model provides a bound on the crosstalk effects. QuSpin provides a general guideline for measuring internal sensor offsets \cite{noauthor_qzfm_nodate}, but have not published any measurement results. 

\revone{In the wider measurement community, the QZFM has seen multi-disciplinary use in studies ranging from neutron experiments such as the TUCAN experiment at TRIUMF and the nEDM2 experiment at Paul Scherrer Institute \cite{abel_large_2023, ayres_very_2022}; Low-field nuclear magnetic resonance spectroscopy \cite{put_zero-_2021}; Bio-magnetism studies both for human \cite{lin_using_2019, boto_measuring_2021}and plant \cite{fabricant_action_2021}; and Brain-Computer Interfacing \cite{guger_optically_2021}. Yet despite its popularity, only piecemeal discussions exist in the published literature regarding QZFM characterization.} The present study outlines sets of procedures used by the TUCAN collaboration in calibrating the QZFMs. These procedures are designed to be easily reproducible in most laboratories with access to magnetic shielding capable of suppressing the environmental field to below \SI{50}{nT}, the working regime of the QZFM. The results of the three characterization studies are presented and their significance is discussed. 

\section{Background}
\label{Background}

At the heart of a QZFM is a circularly polarized \SI{795}{nm} laser resonant with the D1 transition of \ce{^{87}Rb}. Internal optics reflects the light into a \numproduct{3 x 3 x 3} mm cell containing \ce{^{87}Rb} vapor heated to approximately \SI{150}{\celsius}. \cite{shah_fully_2018} The laser establishes a magnetically sensitive state in the Rubidium via optical pumping, wherein the vapor's opacity to the laser becomes a function of the magnetic field strength perpendicular to the beam direction. The maximum amount of transmission occurs at zero field when the magnetic moment is fully aligned with the laser, while any perpendicular field component causes Larmor precession in the atoms, decreasing the vapor transparency. A second prism redirects the transmitted beam onto a photo-detector, where the intensity of the transmitted light is measured. This intensity is proportional to the magnitude of magnetic fields perpendicular to the laser beam. 

\begin{figure*}
    \centering    \includegraphics[width=\textwidth]{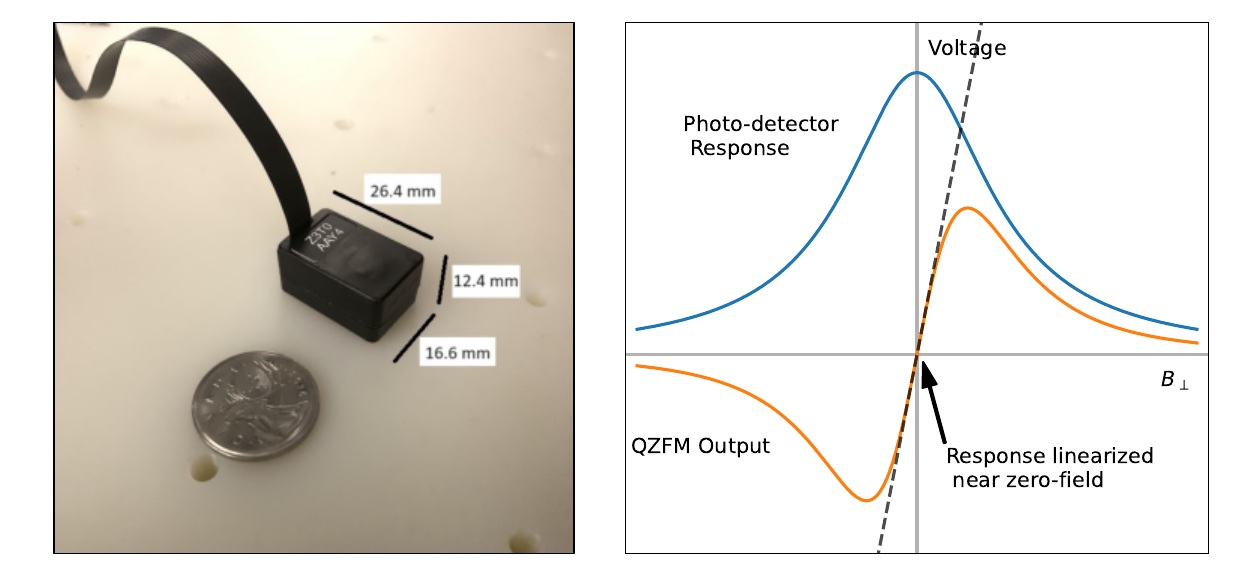}
    \caption{(Left) A 3rd generation QuSpin Zero Field Magnetometer (QZFM), coin for scale. (Right) The QZFM works by shining a laser through a rubidium vapor cell. The transmitted laser intensity exhibits a Lorentzian shape as a function of the magnetic field perpendicular ($B_{\perp}$) to the beam direction. By applying a modulation field and monitoring the demodulated signal, the QZFM output exhibits a dispersion shape, providing the two-fold benefit of linearizing the output near zero and allowing for discrimination of positive and negative fields.}
    \label{fig:QZFMphoto-PDoutput.pdf}
\end{figure*}

The output of the photo-detector as a function of magnetic field strength exhibits a Lorentzian dependence with a zero field resonance. The QZFM further modulates this output with a \SI{923}{Hz} sinusoidal field \cite{tierney_optically_2019}. An onboard phase-sensitive lock-in detector demodulates the signal, producing an odd response about the origin in a dispersive form. This allows the sign of the external field to be determined, as shown in Fig.~\ref{fig:QZFMphoto-PDoutput.pdf}. Through beam splitting the laser and the use of multiple orthogonal modulation fields \cite{boto_triaxial_2022, shah_fully_2018}, simultaneous measurement of the magnetic field along 3 axes can be achieved, making the QZFM a vectorial magnetometer. Finally, as the QZFM is only sensitive near zero-field where the response curve is steep, it can only be used reliably when the background field is below $\sim$\SI{2}{nT}. To relax this stringent requirement, the QZFM sensor head is equipped with 3 sets of compensation coils capable of nullifying up to \SI{50}{nT} in each axis \cite{shah_method_2015}. 

While this device produces field measurements with extremely high precision, it has a number of minor flaws. Firstly, the need for active nulling could introduce systematic offsets in the absolute field reported due to calibration errors in the current source and coil installation. To a lesser extent, a DC offset can also be caused by small remnant fields produced by the materials used in the sensor assembly. Secondly, the QZFM reports voltages proportional to the demodulated signal, shown in Fig.~\ref{fig:QZFMphoto-PDoutput.pdf}. Evidently, the response is non-linear when considered over a sufficiently large range of field magnitudes. Tierney et al \cite{tierney_optically_2019} report \SI{<1}{\percent} deviation from absolute linearity for fields less than \SI{3}{nT_{pp}}, while Boto et al \cite{boto_moving_2018} report \SI{<4}{\percent} in the same range. For magnetic mapping it is highly important to account for this non-linearity because even small field gradients inside a large MSR may lead to several nT of variation in fields throughout the experiment region ($\sim$\SI{2}{m^3} volume). Lastly, the presence of crosstalk can be traced to the modulation fields \revone{driven by Helmholtz coils within the sensor head}. As there can be no shielding about the sensor head, the fringes of the modulation fields can be detected outside the sensor volume. When multiple sensors are placed in close proximity, these fields superimpose, changing the effective modulation field at individual cell centers. QuSpin implements a partial mitigation \revone{through temporal synchronization} \revonedel{by synchronizing the modulation signal} \revone{via sharing the signal from a central `master' sensor} \revonedel{across} \revone{to} all \revone{other} sensors in a system \revone{, referred to as `slaves'.}\cite{shah_fully_2018} This approach removes phase lags between the coils, thus keeping interference consistent and reproducible. However, it does not address the changes in effective modulation field direction and amplitude. \revone{As the gain of the QZFM is proportional to the modulation field, the presence of crosstalk introduces a systematic error.}\cite{tierney_optically_2019}

\section{Methods}
\label{Methods}
\subsection{Data Acquisition}
The magnetic field from the QZFM analog output was sampled with a LabJack T7 DAQ \cite{noauthor_labjack_nodate}. A Python interface was developed for scriptable control and readback of the QZFM sensors \cite{fujimoto_quspin_nodate}.
The interface was used to automate querying the QZFM for quantities including the field produced by the built-in compensation coils, the cell temperature error, and the cell temperature control voltage. The applied compensation field values are necessary for the determination of sensor intrinsic offsets. 

\subsection{Offset}
When a magnetometer reads a DC magnetic field $B$, it may consist of two parts: $B=B_\text{offset}+B_\text{environment}$, where the offset is an undesired systematic error intrinsic to the device and $B_\text{environment}$ is the quantity desired. Since the offset contribution is independent of the sensor orientation, it can be extracted by taking two measurements, $B_{\uparrow}$ and $B_{\downarrow}$, which differ by an inversion of the sensor orientation, whereby the environmental contribution will change signs:

\begin{equation} \label{eq:offset calculation}
B_\text{offset}=\frac{B_{\uparrow}+B_{\downarrow}}{2}.
\end{equation}

\begin{figure*}
    \centering
    \includegraphics[width=\textwidth]{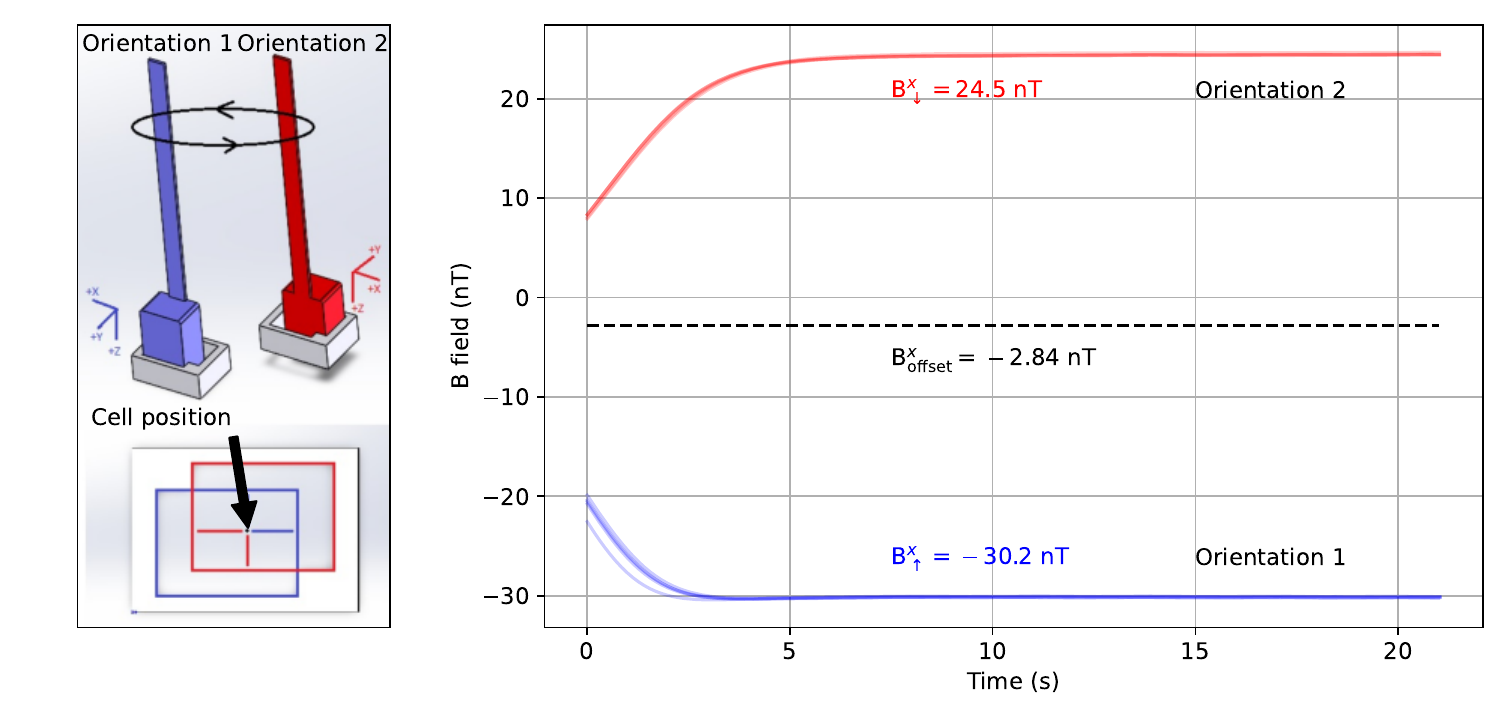}
    \caption{(Left, top) A 3D-printed PLA holder with two overlapping slots. After the flip, the two sensitive axes orthogonal to the rotation axes will be reversed. In each orientation, the sensor is zeroed and the field applied by the compensation coil is monitored. (Left, bottom) Top-down view of the holder, the QZFM placed in any one slot will have its cell fixed in place. (Right) Applied nulling field along $x$ in the two orientations. After the internal control loops for the coils have flat-lined, the field values can be used to compute the offset (dashed line) via Eq.~\ref{eq:offset calculation}.}
    \label{fig:offset_methods}
\end{figure*}

As shown in Fig.~\ref{fig:offset_methods}, 3D printed holders were designed consisting of pairs of partially overlapping slots. The slots permit a consistent 180$^\circ$ rotation of the sensor while constraining the position of the sensor cell. At least two different rotation axes are needed to measure the offset in all three axes. 

The two measurements $B_{\uparrow}$ and $B_{\downarrow}$ are obtained by reading back the zeroing fields applied by the built-in compensation coils. The two orientations are cycled through 5 times each and the calculated offset is taken to be the average of the consecutive pairs. 

\begin{figure*}[t!]
    \centering
    \begin{subfigure}[t]{0.5\textwidth}
        \centering
        \includegraphics[height=2in]{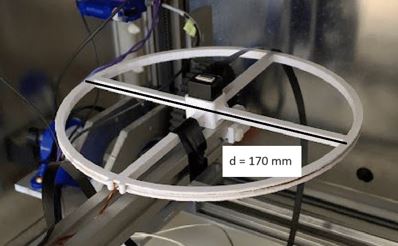}
    \end{subfigure}%
    ~ 
    \begin{subfigure}[t]{0.5\textwidth}
        \centering
        \includegraphics[height=2.3in]{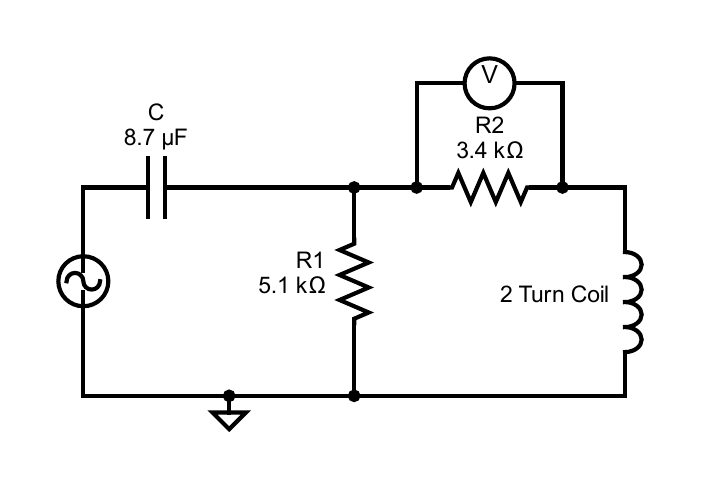}
    \end{subfigure}
    \caption{Setup inside an MSR to assess response linearity. (Left) A coil whose major axis is aligned with the z-axis of the QZFM was used to produce a background oscillation; one coil was printed corresponding to each sensitive axis. (Right) The circuit consists of a first-order high-pass filter to block DC offsets and a voltmeter that measures the voltage across $R_2$ to monitor the current in the loop.}
\label{fig:Linearity-methods}
\end{figure*}

\subsection{Response Linearity}

A Rigol DG1032Z Waveform generator was used to
produce a \SI{35}{Hz} sinusoidal signal into the circuit shown in Fig.~\ref{fig:Linearity-methods}. \revone{The signal serves as a reproducible control excitation. We measure the amplitude of this signal to isolate the QZFM performance from any drift in the background field.} The circuit filters the signal with a first-order high-pass filter to remove any DC offset. The \revonedel{voltage} \revone{current} amplitude of the filtered signal was monitored across a second resistor in series with the coil using a Keithley DMM6500 digital multimeter. The signal was applied as an external oscillating field on the QZFM through the 2 turn 170 mm diameter coil centered on sensor housing, also shown in Fig.~\ref{fig:Linearity-methods}. The driving amplitude was varied between 0 and \SI{18}{Vpp} in 200 discrete steps. At each step the response from the QZFM was measured for 5 seconds \revonedel{and the amplitude was averaged.} \revone{and the average amplitude was obtained by first roughly locating all extremas with a peak-finding algorithm and then using a fitted fourth-order polynomial to obtain the peak values to minimize the influence of higher frequency noise.} This yields the QZFM response as a function of driving voltage, independent of the background field.

\subsection{Multisensor Crosstalk}
QZFM crosstalk arises from the leakage of the modulation field outside of the sensor volume. To assess the presence of this modulation signal, a Stefan Mayer Fluxmaster \cite{noauthor_stefan_nodate} was placed in one-centimeter intervals from a QZFM, starting with the two sensors directly touching. The Stefan Mayer sensor was sampled at \SI{1}{kHz}, aliasing the \SI{923}{Hz} modulation field to appear in the spectrum at \SI{77}{Hz}. The amplitude of the \SI{77}{Hz} component was monitored as a function of the sensor separation distance. 

To further assess the effect of crosstalk, two QZFMs were placed in proximity. The two sensors were set up in the ``master-slave" configuration whereby the modulation signal from the master sensor is shared with the slave sensor. Thus the \revone{temporally synchronized} second sensor acts as an additive perturbation. The field reading from the primary sensor is streamed while varying the separation distance. The same coils used in the linearity study drive a steady background sinusoidal signal at \SI{35}{Hz}. Under the influence of this perturbation, the net modulation field amplitude and direction experienced by the primary sensor are changed, causing an effective gain change in the response.  The shift in the response amplitude referenced at the excitation frequency is monitored. 

To quantify the gain change, we define the fractional change in amplitude relative to when the sensors are directly adjacent as 
\begin{equation}
\label{eq:rho}
    \rho(d) = \frac{s(d)}{s(d_0)},
\end{equation}
where $d$ is the separation distance, $s(d)$ is the crosstalk-induced difference in the field amplitude at distance $d$, and $d_0$ is the minimum possible distance given the integrated sensor housing.

\section{Results and Discussion}
\label{Results}
\subsection{Offset}
\label{Offset}

\begin{table*}
\centering
\begin{tabular}{|c | c c c |} 
 \hline \cline{2-4}
 & \multicolumn{3}{|c|}{QZFM AAL9}\\ 
 \hline
 Experiment Date & x [nT] & y [nT] & z [nT] \\ [0.5ex] 
 \hline 
 \textbf{May 2023} & $-2.79\pm 0.15_{sys} \pm 0.03_{stat}$ & $-0.31\pm 0.15_{sys}\pm0.06_{stat}$ & $3.33\pm 0.15_{sys}\pm0.04_{stat}$ \\ 
\textbf{Feb 2024} & $-2.88\pm 0.29_{sys} \pm0.20_{stat}$ & $-0.29\pm 0.29_{sys} \pm 0.06_{stat}$ & $3.23\pm 0.01_{sys} \pm 0.14_{stat}$ \\
 \textbf{May 2024}  & $-2.84\pm 0.15_{sys} \pm0.04_{stat}$ & $-0.09\pm 0.10_{sys} \pm 0.03_{stat}$ & $3.63\pm 0.17_{sys} \pm 0.12_{stat}$  \\
 \textbf{Jun 2024}  & $-2.79\pm 0.15_{sys} \pm0.08_{stat}$ & $-0.29\pm 0.10_{sys} \pm 0.15_{stat}$ & $3.42\pm 0.25_{sys} \pm 0.06_{stat}$ \\[1ex] 
 \hline
\end{tabular}
\caption{Measured offsets for QZFM serial number AAL9 from four sessions spanning about one year. The real DC field is obtained after subtracting the offsets.}
\label{table:1}
\end{table*}

\begin{table*}
\centering
\begin{tabular}{|c | c c c |} 
 \hline \cline{2-4}
 & \multicolumn{3}{|c|}{QZFM AAY4} \\ 
 \hline
 Experiment Date & x [nT] & y [nT] & z [nT] \\ [0.5ex] 
 \hline 
 \textbf{May 2023} & / & / & / \\ 
\textbf{Feb 2024} & / & / & /\\
 \textbf{May 2024}  & $-2.10\pm 0.12_{sys} \pm0.05_{stat}$ & $-0.08\pm 0.10_{sys} \pm0.08_{stat}$ & $3.78\pm 0.29_{sys} \pm0.22_{stat}$\\
 \textbf{Jun 2024}  & $-2.27\pm 0.15_{sys} \pm0.08_{stat}$ & $-0.06\pm 0.10_{sys} \pm0.06_{stat}$ & $3.91\pm 0.29_{sys} \pm0.25_{stat}$\\[1ex] 
 \hline
\end{tabular}
\caption{Measured offsets for QZFM serial number AAY4 from two sessions spanning two months. The real DC field is obtained after subtracting the offsets.}
\label{table:2}
\end{table*}

In Fig.~\ref{fig:offset_methods} we plot the time evolution of the field applied by the x-compensation coils for 5 trials in each orientation. In each trial the QZFM was given the command to null the field for $\sim$20 seconds continuously. In the first few seconds, the control feedback of the coils causes the generated field to converge to a steady-state value cancelling the field external to the cell \cite{shah_method_2015}. The coils are considered stable once the applied field fluctuations has dropped below \SI{100}{pT/s}. The presence of an intrinsic offset is readily apparent, as the stabilized field readings in the two orientations clearly do not match. 

The 3D printed slots were made to have dimensions \SI{0.2}{mm} larger than the QZFM housing. In principle, this tolerance could result in misalignment up to $\delta=\SI{0.4}{\degree}$ from the ideal orientation, introducing a systematic error proportional to the magnitude of the environment field. The measurements were taken inside of a two-layer MSR but as the shielding was not systematically demagnetized prior to each measurement, a remnant field $B_0$ of  $\sim$\SI{30}{nT} persisted along each axis. The associated systematic error in the offset can be estimated as ${\sigma_\text{sys} = \pm B_0\sin{(\delta)}/\sqrt{2}}$. 

The calculated offsets for the two QZFMs are given in Tables \ref{table:1} and \ref{table:2}. Four measurement campaigns were conducted over more than one calendar year to understand the time variability of these quantities. One of the QZFMs (AAY4) only entered inventory after the first two measurement dates. The magnitude of the offsets across both sensors are less than \SI{4}{nT}, though the y offsets are about one order of magnitude smaller than the others. Across the sessions, the offsets in all three sensitive directions remained stable within \SI{1}{nT}, but statistically significant variations are nonetheless observed. While it does not necessarily follow that this represents the maximum possible variation in a year and a higher frequency of measurement could be beneficial, the stable sub-nT drifts corroborates QuSpin's expectations of stability in the time scale of months \cite{noauthor_qzfm_nodate}. The offset likely stems from calibration errors involving the internal DC current source and nulling coils. Assembly imperfections and magnetization of construction elements could also contribute to the offset. 

Measured DC fields can be corrected by subtracting the offset to obtain a better estimate of the environment field. Previous studies  have also utilized QZFMs for MSR mapping and could benefit from a more accurate field reading.\cite{holmes_bi-planar_2018} This is especially important for assessing ultra-quiet environments where the target field is in the sub-\unit{nT} regime. Knowledge of the offset values are also necessary for calibration of active compensation systems \cite{abel_large_2023}, where a magnetic field is minimized when the QZFM approaches the offset values. 

\subsection{Response Linearity}
\label{Linearity}
\begin{figure}
    \centering
    \includegraphics[width=\textwidth/2]{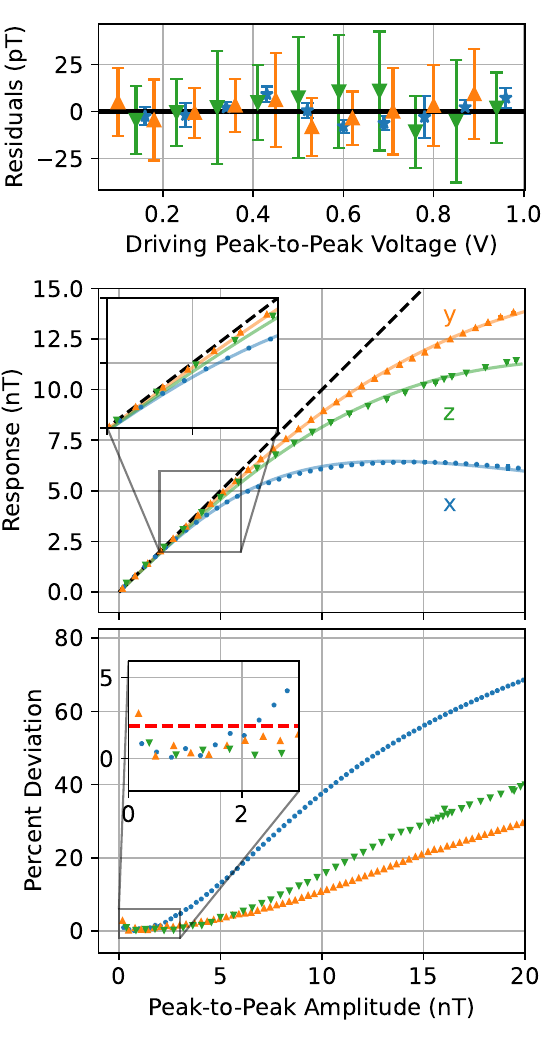}
    \caption{(Top) Residual of the linear curve-fit in the range $V_{\text{pp}}<$ \SI{1}{V}, used in the conversion of applied voltage to field. The QZFM is highly linear in this regime. (Center) Measured response curve to varying sinusoidal field amplitudes. The response is highly non-linear when considered over a \SI{20}{nT_{pp}} range. The inset focuses on the range where non-linearity begins to be significant. The three curves are fitted to Eq.~\ref{eq:inverse_correction_function}, only every third measured point is shown to prevent visual clutter.  (Bottom) Percent deviation from linearity. The inset shows a detailed view of the near-zero field regime. The response remains within 2 percent linear when the field is below \SI{2}{nT_{pp}}.}
    \label{fig:linearity_results}
\end{figure}

In Fig.~\ref{fig:linearity_results} the results of the response linearity study are summarized. The response curve as a function of voltage is dependent on the geometry of the coil setup. A geometrically independent function of the applied field was obtained by measuring the voltage at the resistor $R_2$ (see Fig.~\ref{fig:Linearity-methods}) and converting to nT using the slope of the QZFM response at small amplitudes. Below \SI{1}{V_{pp}} the response is highly linear, as shown in the top of Fig.~\ref{fig:linearity_results}. The converted response curves for one of the QZFMs are plotted in the center graph of Fig.~\ref{fig:linearity_results}. In all three axes, non-linearity can be observed. However, the severity of non-linearity is different, with the most rapid deviation from linearity in the x-axis. \revonedel{This was not the case for both sensors tested: of the second QZFM the z-axis was found to be linear over the smallest range.} The severity of non-linearity can be quantified by considering the percent deviation from linearity, plotted in the bottom graph of Fig.~\ref{fig:linearity_results}. The red dashed line in the inset marks a 2 percent deviation from linearity. From this, it is concluded the QZFM being tested is linear within 2 percent for fields within \SI{2}{nT_{pp}}. \revonedel{A similar benchmark was reached for the second QZFM.} \revone{The same experiment was conducted on the second TUCAN QZFM. While overall the sensor remained within 2 percent linear in a comparable range (\textless \SI{3}{nT_{pp}}), the z-axis was found to be linear over the smallest range rather than x.}

The present results are generally in agreement with previous studies. QuSpin \cite{shah_fully_2018} measured an response linearity to within 1 percent when the signal amplitude is less than \SI{2}{nT_{pp}} while Boto et al\cite{boto_moving_2018} and Tierney et al \cite{tierney_optically_2019} respectively found the signal to be within 4 percent and 1 percent of linearity with fields below \SI{3}{nT_{pp}}. Note in the Boto et al study the methodology was different in that the DC offset was varied instead of the oscillation amplitude. \revone{While all results exhibit a comparable degree of linearity, small variations nonetheless exist. To our knowledge, the present study contains the first demonstration of differing degrees of non-linearity among the QZFM axes. The sample size remains too small for a statistically confident bound on all QZFMs.}

The practical application of the obtained results is the effective extension of the operation range. The response follows the shape of a dispersion curve, as is particularly evident for the response of the x-axis. This dispersion can be described to leading order with the form \cite{tierney_optically_2019}
\begin{equation}
\label{eq:forward_equation}
    B_\text{meas}=A_0\frac{\gamma B_\text{true} \tau}{1+(\gamma B_\text{true} \tau)^2},
\end{equation}
where $A_0$ is an amplitude scaling constant, $\gamma$ is the gyromagnetic ratio of \ce{^{87}Rb}, roughly \SI{7 }{Hz/nT}, $\tau$ is the effective relaxation time of polarized vapor, $B_{\text{meas}}$ the measured response, and $B_{\text{true}}$ the true background field. Here $A_0$ and $\tau$ are the free fitting parameters while $\gamma$ is kept fixed throughout. The inverse function of Eq.~\ref{eq:forward_equation} can be used to correct the measured field values, up to the field value where the inverse function ceases to be single-valued: 
\begin{equation}
\label{eq:inverse_correction_function}
    B_\text{true} = \frac{A_0-\sqrt{A_0^2-4 B_\text{meas}^2}}{2\gamma B_\text{meas} \tau}.
\end{equation}
The response curves fitted to Eq.~\ref{eq:inverse_correction_function} are shown in Fig.~\ref{fig:linearity_results}. The goodness of fit is assessed quantitatively using the reduced chi-squared statistic ${\chi^2_{\nu}}$, yielding for the three axes ${\chi^2_{\nu, x}=73.7}$, ${\chi^2_{\nu, y}=10.1}$, and ${\chi^2_{\nu, z}=51.4}$ respectively. These rather high chi-square values are likely due to the fit function, Eq.~\ref{eq:forward_equation}, being only the leading term of the full solution. This is supported by the trend ${\chi^2_{\nu, y}<\chi^2_{\nu, z}<\chi^2_{\nu, x}}$, with the $x$ response curve already past the maximum sensitive point while that of $z$ and $y$ still to reach this point in the measured range.

Without the correction function, the QZFM is accurate only if the fields are within \SI{1}{nT}. Exceeding which, the QZFM needs to be re-zeroed, a process that increases the temporal cost of many experiments and makes automation more difficult. For biomagnetic studies this often constraints the subject's allowed movement \cite{sander_avoiding_2021}. In addition, having to re-zero between two measurements introduces further uncertainties associated with the nulling coils. With the correction function, the acceptable field range is increased by at least an order of magnitude. 

\subsection{Multisensor Crosstalk}
\label{Crosstalk}
\begin{figure}
    \centering
    \includegraphics[width=\textwidth/2]{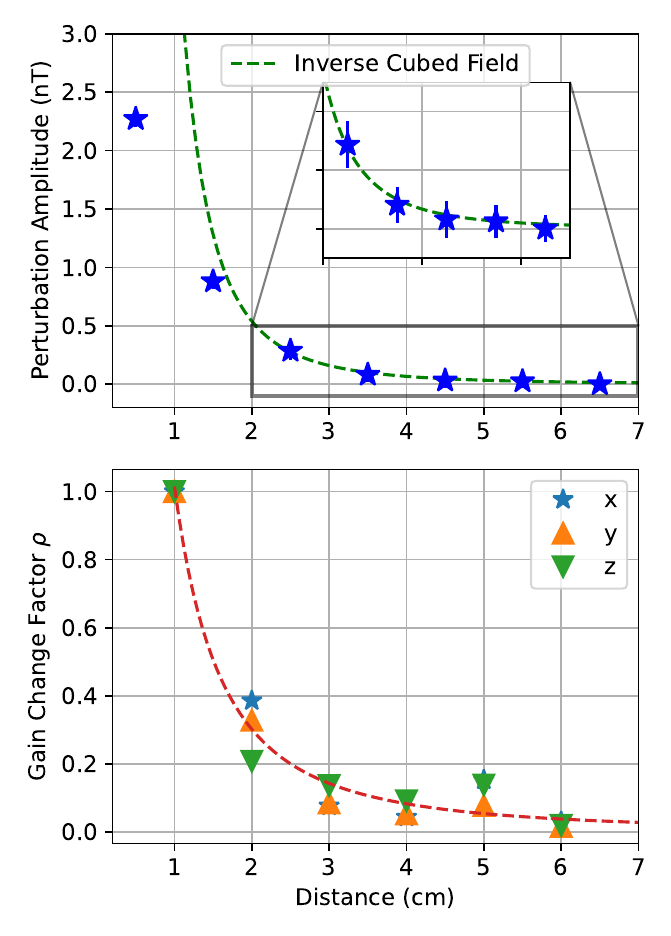}
    \caption{(Top) Amplitude of the modulation signal seen outside the sensor volume measured using a Stefan-Mayer fluxgate. This signal leakage is the cause of perturbation on other sensors. (Bottom) Gain change on a QZFM as quantified by Eq.~\ref{eq:rho} due to presence of a second QZFM. This is a direct measurement of the crosstalk effect. Both measurements of crosstalk diminishes rapidly with distance, and are suppressed below 3 percent with 6 cm of separation between two sensors.}
    \label{fig:crosstalk_results}
\end{figure}

The measured amplitude of the leakage modulation field by the Stefan Mayer fluxgate can be seen in the top graph of Fig.~\ref{fig:crosstalk_results}. Directly outside the integrated sensor housing the amplitude is about \SI{2.5}{nT} and decreases thereafter. After the initial few centimeters, the field decays proportional to the inverse cube of the distance, consistent with the far-field approximation of a loop of current. At the center of the vapor cell the amplitude of the modulation field is \SI{60}{nT} \cite{tierney_optically_2019}, from which it can be deduced that even directly outside the sensor housing, the amplitude has attenuated to less than 4 percent of the center strength. This result is consistent with those obtained by Boto et al \cite{boto_moving_2018} where the measured field after attenuation was in the range between 1 and 3 percent for sensors placed with similar separations as the present study. However, the Boto study used bi-axial first generation QZFMs compared to the tri-axial third generations under present discussion. Furthermore, the sensors were placed on a helmet, with sensors at angles to one another. 

The gain change due to crosstalk was measured directly with a second QZFM as shown in the bottom graph of Fig.~\ref{fig:crosstalk_results}. By definition, the gain change factor $\rho$ (Eq.~\ref{eq:rho}) is unity at the minimum possible distance between two QZFMs. The amount of gain change then drops as the leakage modulation field from the perturbing sensor becomes negligible.

\begin{figure}
    \centering
    \includegraphics[width=\textwidth/2]{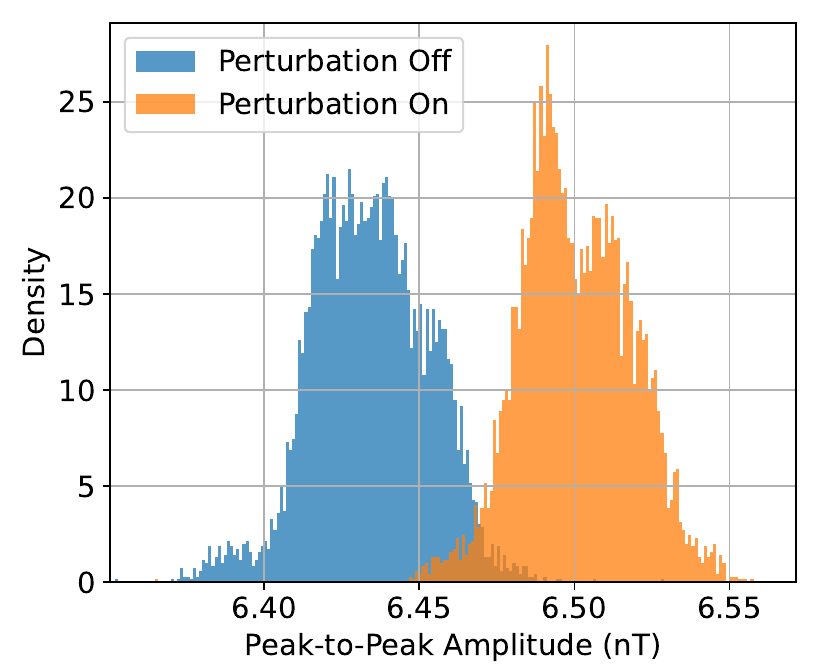}
    \caption{Two QZFMs are placed directly adjacent. A constant background sinusoidal signal is present and the amplitude is reported by the primary sensor. The histograms show the amplitude before and after perturbation. A noticeable shift in the measured amplitude appears when the perturbing sensor is powered on.}
    \label{fig:crosstalk_methods}
\end{figure}

From a practical perspective, the induced gain change appears to be a minor effect for almost all scenarios. In terms of absolute magnitudes, for a driving field of roughly \SI{6.4}{nT} shown in Fig.~\ref{fig:crosstalk_methods}, where the perturbing and primary sensor are in the worst case configuration (directly touching), the field reading change was only on the order of about \SI{100}{pT}. 

We conclude that with a separation above \SI{6}{cm}, crosstalk can be safely neglected as a dominating error contribution. The benefits of operating two or more sensors simultaneously in close proximity are myriad. For the nEDM experiment, getting two sensors independently monitoring the magnetic field will allow for compensation against sensor drift as well as increase confidence in the sensor readings \revone{by checking the correlation between sensor readings. Validation of planned internal coil-generated fields and the check for permanent magnetization of apparatus internal to the MSR will also depend on having two sensors in a gradiometer configuration.} Beyond this, many biomagnetic applications such as magnetoencephalography and magnetocardiography not only benefit, but in fact require sensor arrays consisting up to hundreds of OPMs to gather enough field information for signal processing.

\section{Conclusion}
Two tri-axial 3rd generation QuSpin Zero Field Magnetometers (QZFMs) were characterized in terms of their intrinsic offset, response linearity, and crosstalk in preparation for the TUCAN nEDM experiment. We trace the source of the three sensor defects from the fundamental principles of  Rubidium zero-field resonance magnetometry and present the procedures of characterization. The offset in all three axes were determined to be roughly in the range \SI{\pm3}{nT} and remained stable to within \SI{1}{nT} over one year. The response is shown to be linear within 2 percent up to \SI{2}{nT} peak-to-peak and beyond this the non-linearity is mapped to high precision and can be well described by Eq.~\ref{eq:forward_equation}, \revonedel{extending the effective measurement range from \num{2} \si{nT_{pp}} to above \num{20} \si{nT_{pp}}.} \revone{extending the effective measurement range from \SI{2}{nT_{pp}} to at least \SI{10}{nT_{pp}} for all axes, with some axes to above \SI{20}{nT_{pp}}.} Multi-sensor crosstalk is present but largely insignificant, being largely negligible for inter-sensor distances above \SI{6}{cm}. The same protocols outlined in this paper conducted on a statistically significant population of QZFMs may provide directly applicable quantities such as bounds on offsets and region of linearity. Furthermore, the study of multi-sensor crosstalk in this paper is for the simple case of only two sensors in parallel. A more complex array geometry could be considered for specific applications. In any case the results from the present work will help enable highly accurate and precise measurements of residual magnetic fields inside the large magnetically shielded room for the nEDM experiment at TRIUMF.

\section*{Data Availability}
The data that support the findings of this research are publically available at the following URL/DOI: \url{https://zenodo.org/records/13231342}

\section*{Conflicts of Interest}
There are no conflicts of interest in the scope of this study.

\ack
The authors would like to acknowledge the members of the TUCAN collaboration for their support. Funding for this work and the TUCAN collaboration was supported by the Canada Foundation for Innovation; the Canada Research Chair program; the Natural Sciences and Engineering Research Council of Canada (NSERC) SAPPJ-2016-00024, SAPPJ-2019-00031 and SAPPJ-2023-00029; Japan Society for the Promotion of Science (JSPS) KAKENHI grants 18H05230 and 20KK0069; JSPS KENHI grants 17K14307, 19K23442, 20K14487, 21K13940, 22H01236; RCNP CORENet.

\section*{References}
\bibliography{main}

\end{document}